\documentclass[twocolumn,superscriptaddress,prl,showpacs]{revtex4-1}
\usepackage{mathptmx}
\usepackage[T1]{fontenc}
\usepackage[latin9]{inputenc}
\usepackage{float}
\usepackage{amssymb}
\usepackage{graphicx}
\usepackage{esint}
\usepackage{color}
\usepackage{hyperref}
\usepackage[normalem]{ulem}
\usepackage{pdfpages} 
\usepackage{pgffor} 

\makeatletter
\AtBeginDocument{\let\LS@rot\@undefined}
\makeatother

\def\supplementfilename{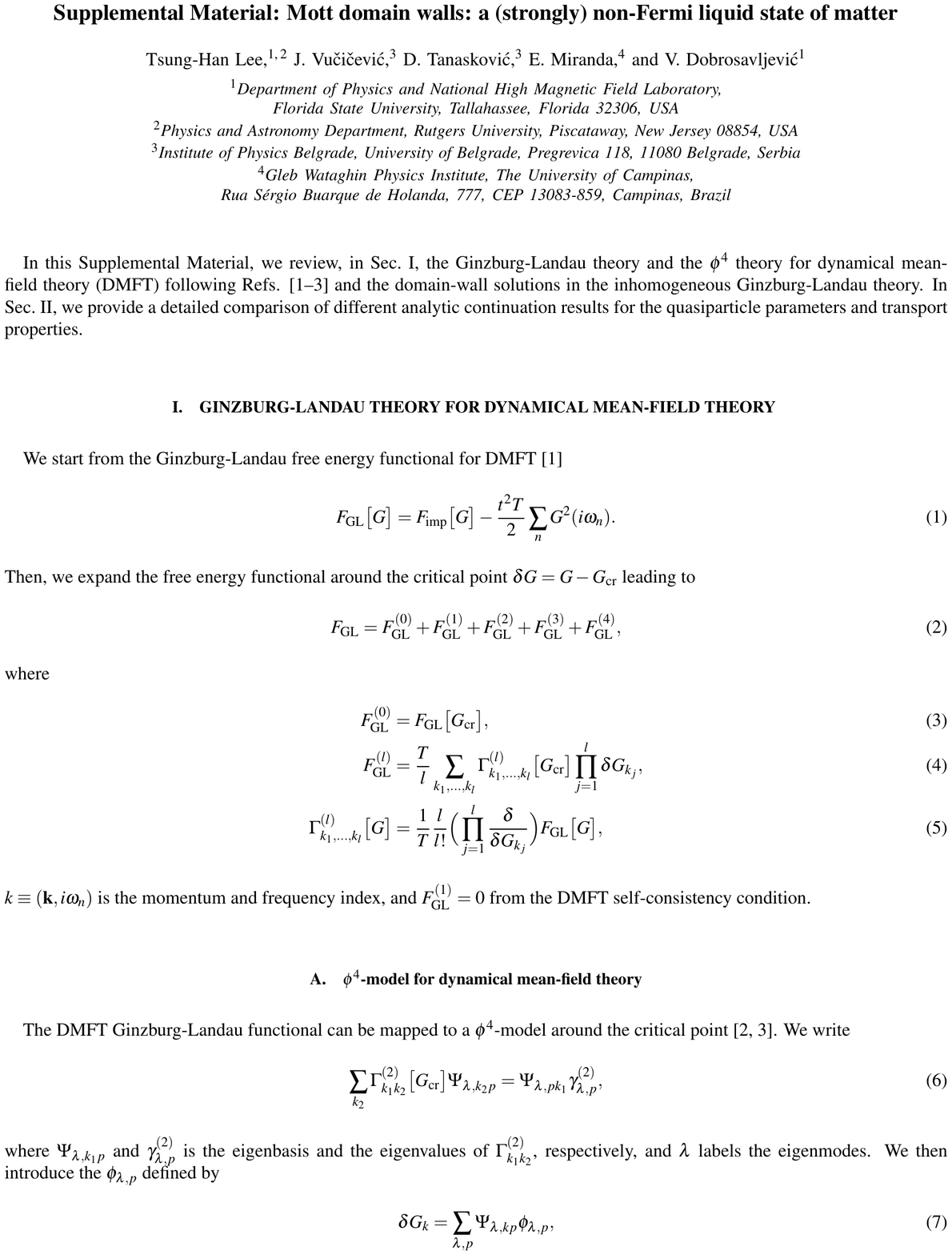}

\pdfximage{\supplementfilename}
\def\numbersupplementpages{\the\pdflastximagepages}

\newif\ifarXiv
\arXivtrue 

\makeatletter

\renewcommand\[{\begin{equation}}
\renewcommand\]{\end{equation}} 

\@ifundefined{showcaptionsetup}{}{%
 \PassOptionsToPackage{caption=false}{subfig}}
\usepackage[caption=false]{subfig}
\makeatother

\begin{document}
\title{Mott domain walls: a (strongly) non-Fermi liquid state of matter}
\author{Tsung-Han Lee}
\affiliation{Department of Physics and National High Magnetic Field Laboratory, Florida State University, Tallahassee, Florida 32306, USA}
\affiliation{Physics and Astronomy Department, Rutgers University, Piscataway,
New Jersey 08854, USA}

\author{J. Vu\v{c}i\v{c}evi\'{c}}
\affiliation{Institute of Physics Belgrade, University of Belgrade, Pregrevica 118, 11080 Belgrade, Serbia}

\author{D. Tanaskovi\'{c}}
\affiliation{Institute of Physics Belgrade, University of Belgrade, Pregrevica 118,
11080 Belgrade, Serbia}
\author{E. Miranda}
\affiliation{Gleb Wataghin Physics Institute, The University of Campinas, Rua S\'ergio Buarque de Holanda, 777, CEP 13083-859, Campinas, Brazil}
\author{V. Dobrosavljevi\'c}
\affiliation{Department of Physics and National High Magnetic Field Laboratory, Florida State University, Tallahassee, Florida 32306, USA}

\begin{abstract}
Most Mott systems display a low-temperature phase coexistence region around the metal-insulator transition. The domain walls separating the respective phases have very recently been observed both in simulations and in experiments, displaying unusual properties. First, they often cover a significant volume fraction, thus cannot be neglected. Second, they neither resemble a typical metal nor a standard insulator, displaying unfamiliar temperature dependence of (local) transport properties. Here we take a closer look at such {\em domain wall matter} by examining an appropriate  {\em unstable} solution of the Hubbard model. We show that transport in this regime is dominated by the emergence of ``resilient quasiparticles'' displaying strong non-Fermi liquid features, reflecting the quantum-critical fluctuations in the vicinity of the Mott point. 
\end{abstract}

\maketitle

{\em Introduction.---} The Mott metal-insulator transition \cite{Mott1990,Imada} remains a subject of much controversy and debate, with disagreement even concerning the physical mechanism \cite{dobrosavljevic2012conductor} that dominates its vicinity. One popular viewpoint \cite{senthil2008prb} regards it as a (strictly) second-order phase transition at $T=0$, where the dominant degrees of freedom are the inter-site spin singlets (e.g. the "spinon" excitations) arising in the vicinity of the Mott insulating state. A complementary Dynamical Mean-Field Theory (DMFT) perspective \cite{A.Georges1996} builds on the seminal ideas of Hubbard and Mott, focusing on local Kondo-like processes that govern the condensation \cite{pustogow2021rise}  of the strongly-correlated Fermi liquid on the metallic side. The latter viewpoint predicts the ``evaporation'' of the electron liquid at the transition bears some analogy to conventional liquid-gas transitions, with a phase coexistence region arising at low temperatures \cite{A.Georges1996}. While many experiments \cite{LimeletteScience2003,kanoda2005nature} indeed reported the predicted first-order transition within the paramagnetic phase, other experiments \cite{furukawa2018natcomm,moire-mott2021} reported behavior consistent with quantum criticality, which sometimes has been interpreted in terms of the former picture \cite{senthil2008prb}. 

Resolving this important issue in the context of real materials is further complicated by the emergence of various magnetic, charge, or orbital orders in the vicinity of the Mott point  \cite{Imada}, which can often mask the basic underlying mechanism. Recent experimental work, however, has successfully identified \cite{kanoda2005prl} a simpler class of model systems, where no broken symmetry phases have been observed anywhere in the phase diagram, providing new information. This situation is best documented in ''spin-liquid'' organic materials \cite{dressel2020advphys}, where careful and precise experiments are starting to paint a clearer picture of the genuine Mott point. Most remarkably, experiments here provided \cite{furukawa2015natphys} clear evidence for quantum critical scaling  \cite{TerletskaPRL} of the resistivity curves at intermediate temperatures, with some evidence for a resistivity jump  at $T < T_c \sim 30K$, consistent with the DMFT prediction of a weekly first-order transition \cite{JaksaWidomPRB}. Still, direct evidence of the phase coexistence has emerged only 
in the recent reports of a colossal enhancement of the dielectric response~\cite{pustogow2021npjQM} and the previous near-field infrared imaging \cite{BasovMITexperiment}.

In parallel with the experimental progress, recent theoretical work provided complementary insight into the nature of the metal-insulator coexistence region \cite{martha2020prb}. Surprisingly ``thick'' domain walls were observed \cite{martha2021prb}, which are likely to play a central role in governing the observable response in experiments. Indeed, the local transport properties of such domain walls were found \cite{martha2020prb,martha2021prb} to display a variety of unusual features, with properties not akin to either those of the conventional metal nor of the insulator. To obtain clear and precise insight into the physical nature of such {\em Domain Wall Matter} (DWM), we present in this paper model calculations within the framework of the DMFT picture. We argue that, similarly to conventional Landau theories for domain walls, the {\em central region} of a domain wall corresponds to a ``saddle point'' (unstable solution) of the spatially uniform DMFT equations, at the top of the free-energy barrier separating the two competing phases~\cite{chaikin_lubensky_1995,supplemental_material}. In dramatic contrast to the conventional Landau theory (e.g. for the Ising model), here the two solutions are {\em not} related by symmetry and display very different physical behavior~\cite{GLDMFT1999,GLDMFT2000,GLDMFT2004}. One is a Fermi liquid metal with coherent quasiparicles (QP) and $T^2$ resistivity, whereas the other is a Mott insulator with completely incoherent activated transport. What should, then, be the physical properties of the unstable solution separating them? Should it resemble more closely a metal or a Mott insulator? What are the thermal properties of transport in this unfamiliar regime? As a matter of fact, it is almost impossible to guess. Previous studies of the unstable solution were restricted only to the near vicinity of the critical end-point ($T\sim T_c$) \cite{TongUnstablePRB,HugoUnstablePRB,Rozenbergdoubleoccupancy}, but they did not provide clear insight as to what happens through the phase coexistence region.

In this paper we present a clear and yet somewhat surprising answer to all these important questions, in a setup which can be considered as a (numerically) exact solution of one of the simplest toy model of strong correlations. Based on a reliable numerical solution of the corresponding DMFT equations we establish that: (1) Transport in DWM assumes ``Resilient QuasiParticle'' (RQP) character \cite{WenhuHFL,DengRQP} throughout the coexistence region; (2) The relevant RQPs display surprising non-Fermi liquid  $T$-dependence of the QP parameters (see below). These results firmly establish that DWM is a qualitatively different form of matter, which we associate with quantum critical fluctuations around the Mott point. 

\begin{figure}[t]   
\centering
\subfloat[]{\includegraphics[clip,scale=0.25]{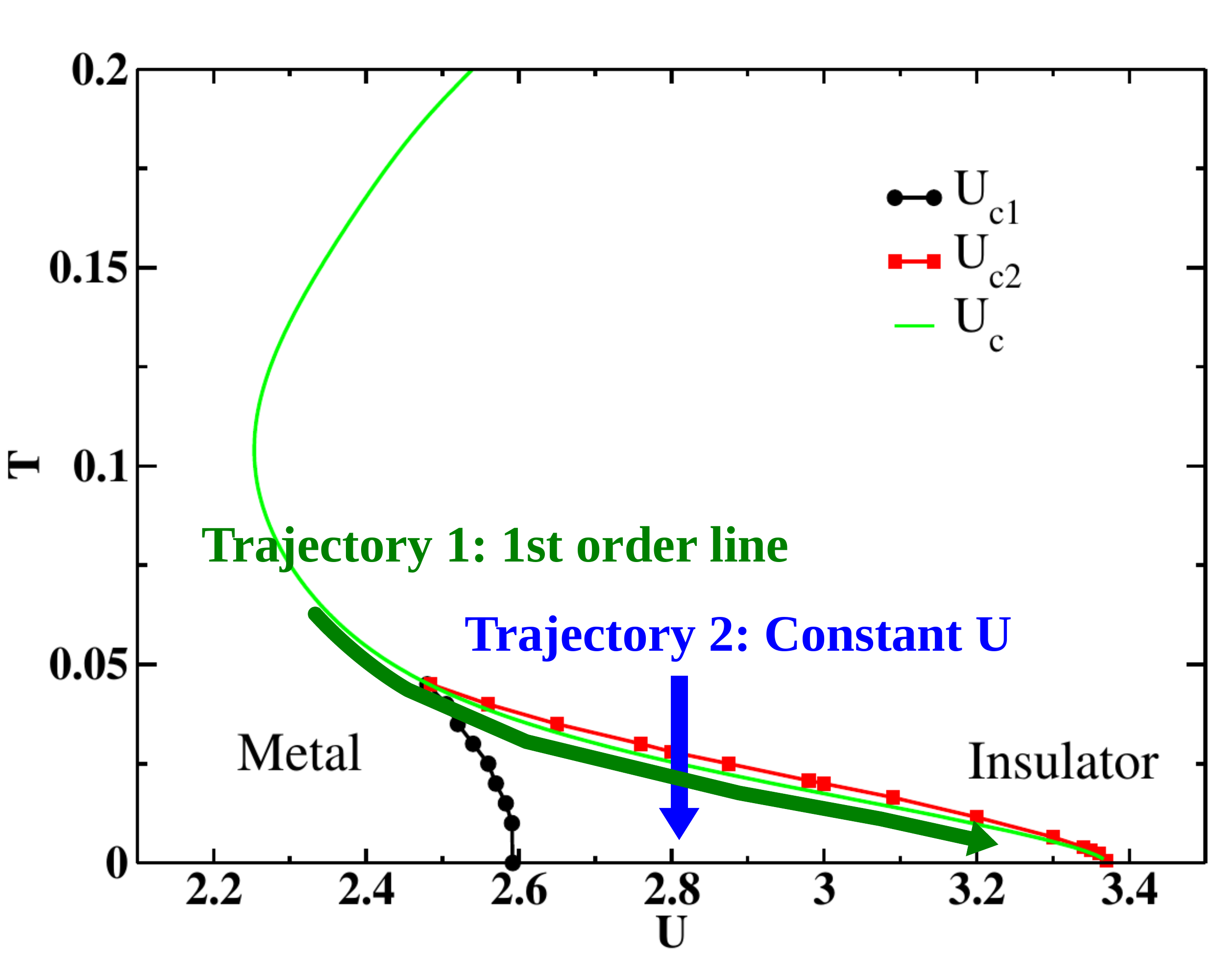}}


\subfloat[]{\includegraphics[clip,scale=0.23]{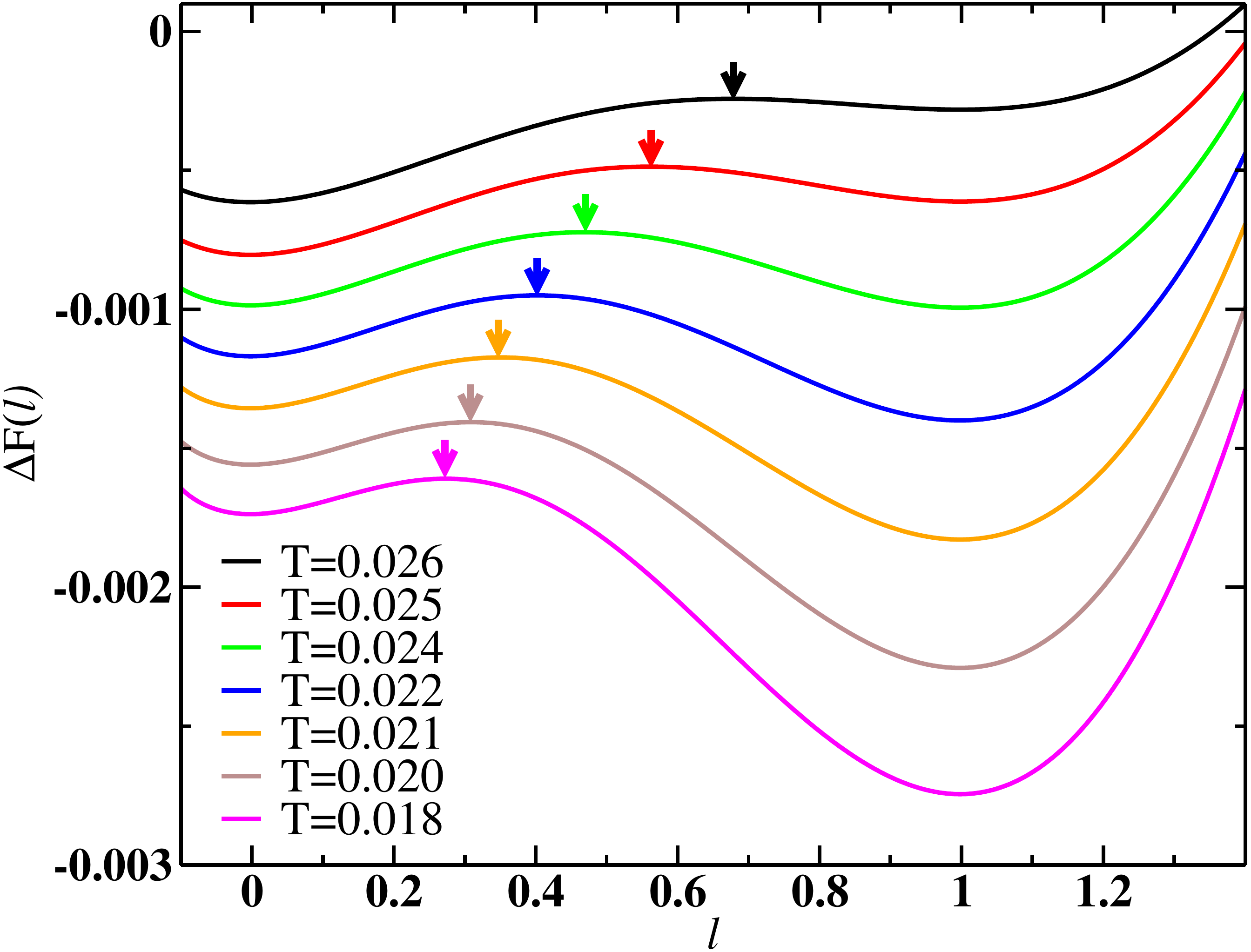}}

\caption{(a) DMFT phase diagram for the half-filled Hubbard model and the trajectories studied (see text).
(b) The evolution of the free energy functional \cite{FEVladPRB} as $T$ varies for the fixed-$U$  trajectory ($U=2.83$). Here \ $l=0$ and $l=1$ correspond to the
insulating and metallic solutions, respectively. The arrows mark the position of the unstable solutions. \label{fig:Free_energy}}
\end{figure}

{\em Model calculations.---} To describe the Mott metal-insulator transition (MIT), while suppressing all forms of magnetic orders, we focus on the ``maximally frustrated Hubbard model'' \cite{A.Georges1996,TerletskaPRL,Dobrosavljevic_Kotliar1993}, given by the Hamiltonian:
\begin{equation}
H=-\sum_{\langle i,j\rangle\sigma} t_{ij}(c_{i\sigma}^{\dagger}c_{j\sigma}+h.c.)+U\sum_{i}n_{i\uparrow}n_{i\downarrow},\label{eq:Hubbard_model}
\end{equation}
where $c_{i\sigma}^{\dagger}$ and $c_{i\sigma}$ are the electron
creation and annihilation operators, $n_{i\sigma}=c_{i\sigma}^{\dagger}c_{i\sigma}$,
$t_{ij} $ are the  hopping elements with zero average and variance $\langle t_{ij}^2 \rangle = t^2 /\sqrt{N}$, and $U$ is the onsite Coulomb
potential. The energy unit is set to the half band width, $D=2t$.
Similarly to the popular SYK model \cite{syk2021},  such an infinite-range model can be exactly solved in the limit where the number of sites
$N\rightarrow \infty$. In this case, this is performed through self-consistently solving an
Anderson impurity model using the DMFT framework \cite{A.Georges1996}. To solve the impurity problem, we utilize well-known continuous
time quantum Monte Carlo (CTQMC) methods as well as iterative perturbation theory
(IPT) as impurity solvers~\cite{GullCTQMC,KajueterIPT}. The analytical continuation to the real-frequency axis is done by
using maximum entropy method (MEM), and the 5th order polynomial fitting
for CTQMC, and the Pad\'{e} approximant for IPT \cite{VidbergPade,JarrellMEM}. The use of an appropriate N-dimensional optimizer \cite{HugoUnstablePRB} is essential for the convergence to the local saddle point of the free energy functional
(the unstable solution). In addition, an appropriate free energy analysis \cite{KotliarLandauTheory,FEVladPRB}
allows us to identify the first-order transition line, as well as the location
of the unstable solution.

The DMFT phase diagram (obtained from IPT), Fig.~\ref{fig:Free_energy} (a), features
a second-order critical end point at $T=T_{c}\sim0.045$. 
 Below $T_{c}$, there emerges a phase coexistence region confined by
two spinodal lines $U_{c1}(T)$ and $U_{c2}(T)$, marking the respective instabilities of the Mott insulator and the metallic solutions. At $T=0$, the first order transition
line merges with the spinodal line $U_{c2}(T)$, and the insulating solution becomes marginally unstable exactly at $T=0$ \cite{FEVladPRB}. Above $T_c$, $U_{c1}(T)$ and $U_{c2}(T)$ merge to form a single Widom line, determined from the minimum of the Landau free energy~\cite{TerletskaPRL}.

\textit{Finding the unstable solution.---} In order to understand
the behavior of all three solutions of our DMFT theory (metal, insulator, and unstable),
we employ the Landau free energy functional method \cite{KotliarLandauTheory,FEVladPRB},
which provides information about the form of the free energy landscape. Within DMFT,
the free energy can be considered as a functional of the local Green's function, $G(i\omega_{n})$,
which for our model assumes the form $F[G]=F_{imp}[G]-t^{2}T\sum_{n}G^{2}(i\omega_{n})$, where $F_{imp}[G]$ is the free energy functional of the associated impurity problem.
When solving the DMFT equations by the standard iteration method, 
one finds convergence  \cite{FEVladPRB} to a given local minimum
of the free energy functional, depending on the initial guess for
$G(i\omega_{n})$. Within the coexistence region,  two stable 
solutions separated by the unstable solution (saddle point) are found. 
To illustrate this, we follow a ``phase space path'' \cite{FEVladPRB} connecting the 
two solutions, which can be parametrized as:
$G(l)=(1-l)G_{ins}(i\omega)+lG_{metal}(i\omega)$
with a parameter $l\in[0,1]$. The corresponding variation of the free energy can be calculated \cite{FEVladPRB}
by evaluating the line integral $\Delta F(l)=t^{2}T\int_{0}^{l}dl'e_{l}\cdot \delta G(G(l'))$,
where $e_{l}=(G_{ins}-G_{met})/|G_{ins}-G_{met}|$ and $\delta G=G_{imp}(G)-G$. Here, the dot product denotes a trace over Matsubara frequencies and $G_{imp}$ is the impurity Green's function dependent on the initial condition of $G$.

The unstable solution we seek exists anywhere within the phase coexistence region. 
To be concrete, however, we examine its evolution following  two specific trajectories: (1) 
along the first order transition line (where the two stable solutions have the same free energy); 
(2) a trajectory where we vary $T$ at constant $U$, Fig. \ref{fig:Free_energy}(a). 
For illustration, in Fig. \ref{fig:Free_energy}(b) we follow Ref.~\cite{FEVladPRB}  
and plot the free energy along trajectory  (2) (constant $U$). 
Here we observe how the unstable solution shifts towards the insulating solution as we lower the temperature and finally merges with the insulating solution at $T=0$. This confirms that the
insulating solution becomes unstable precisely at $T=0$ throughout the phase coexistence region. 

To be able to precisely converge to the desired unstable solution, we should keep in mind that the standard iterative method
(essentially a ``steepest descent'' method) can only find local minima of the free energy, i.e., only the stable solutions. Instead, we use the Broyden method, which can converge to any {\em extremum} of a given functional 
(including saddle-points) \cite{HugoUnstablePRB}, if the initial guess is sufficiently close to the given extremum. Indeed, we 
find that this  method can efficiently converge even to the unstable solution within only a moderate number of iterations. We should stress that 
the unstable solution found in this way is generally not restricted to lie exactly on the ``phase space path'' connecting the two stable solutions, except very close to $T = T_c$. Still, Broyden-converging to the proper unstable solution is greatly facilitated by using the ``phase space path'' to estimate an appropriate initial guess for the root search.

\begin{figure}[t]
\subfloat[CTQMC]{\includegraphics[clip,scale=0.165]{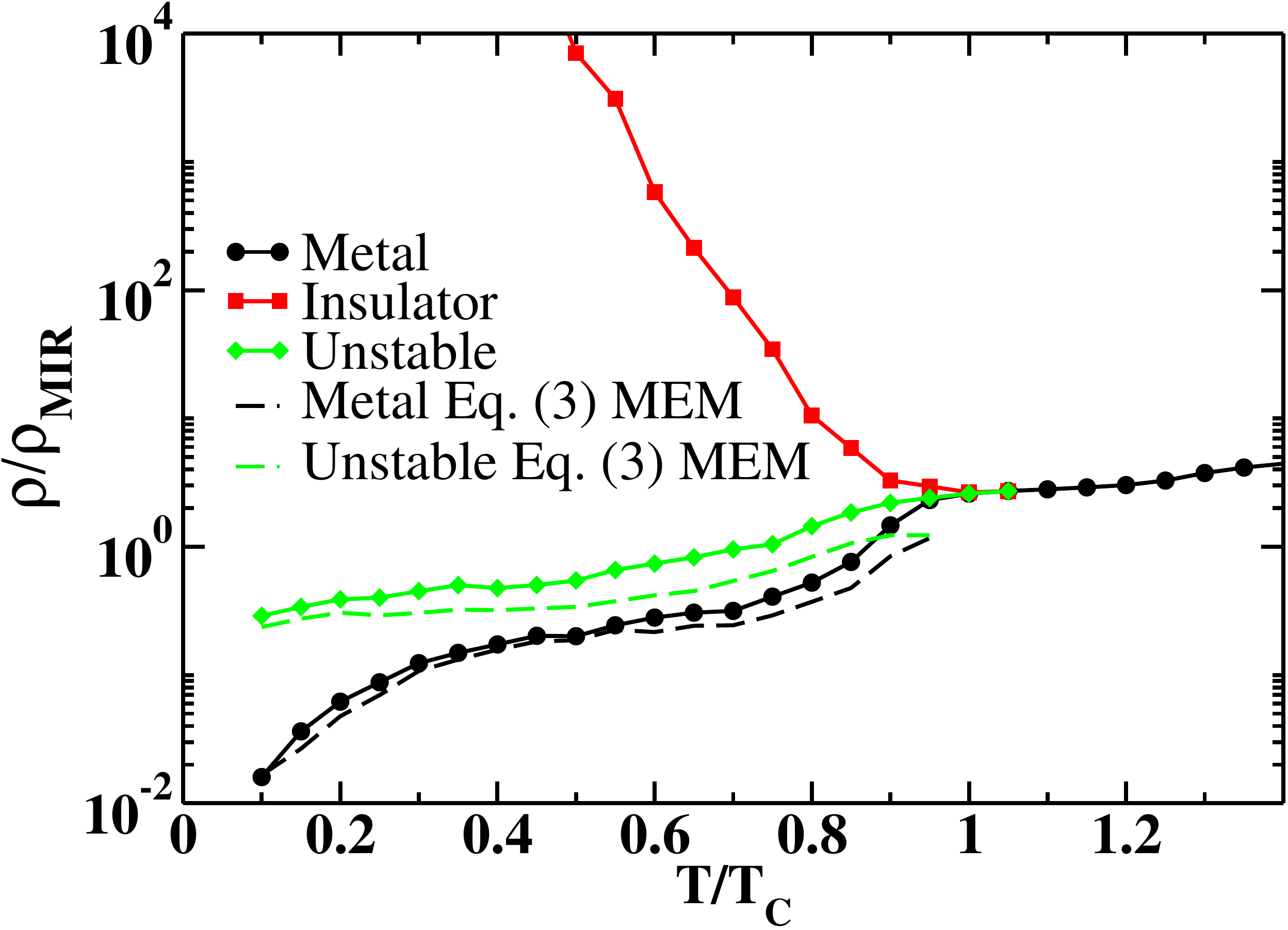}

}\subfloat[IPT]{\includegraphics[clip,scale=0.165]{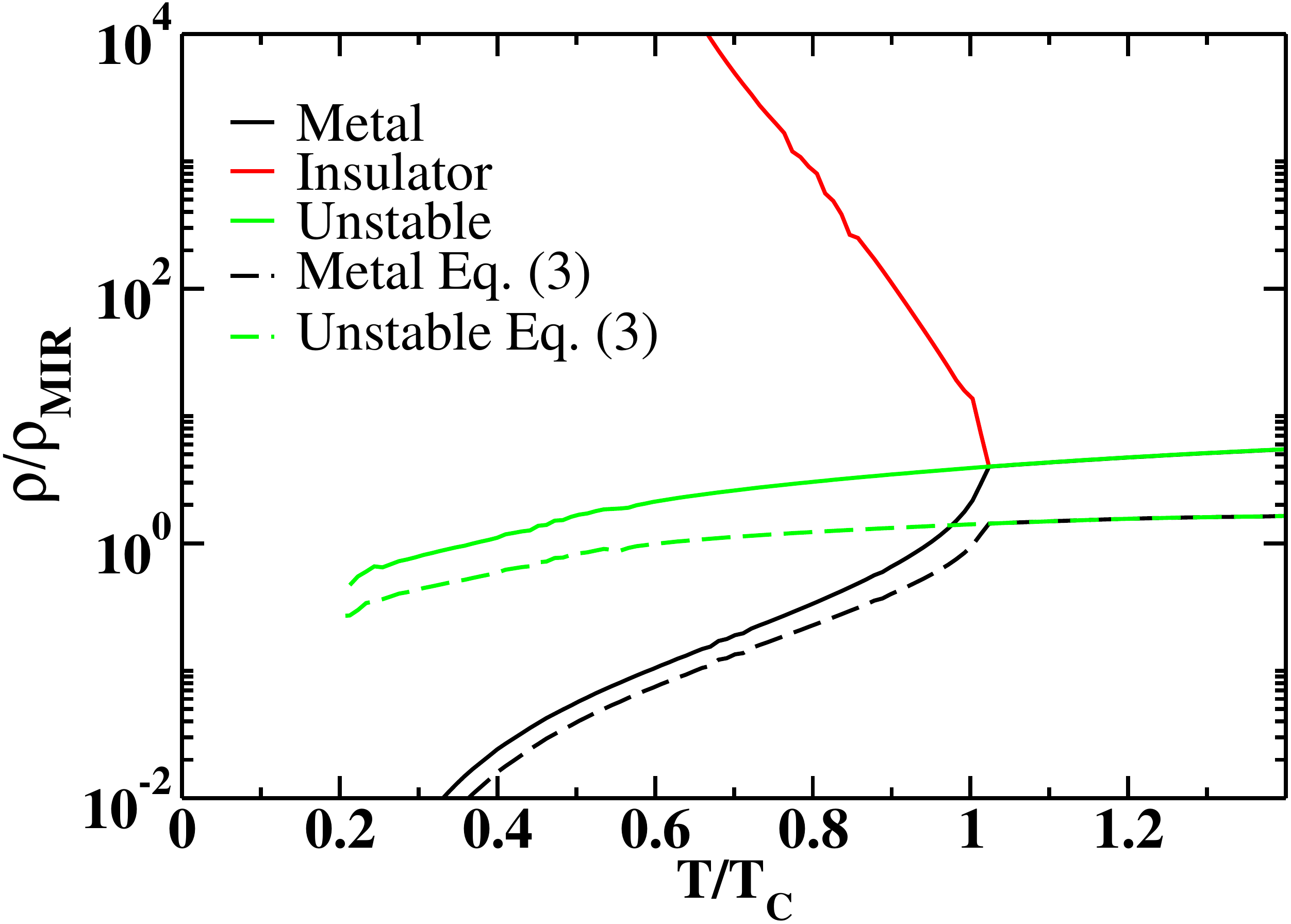}

}


\subfloat[CTQMC]{\includegraphics[clip,scale=0.17]{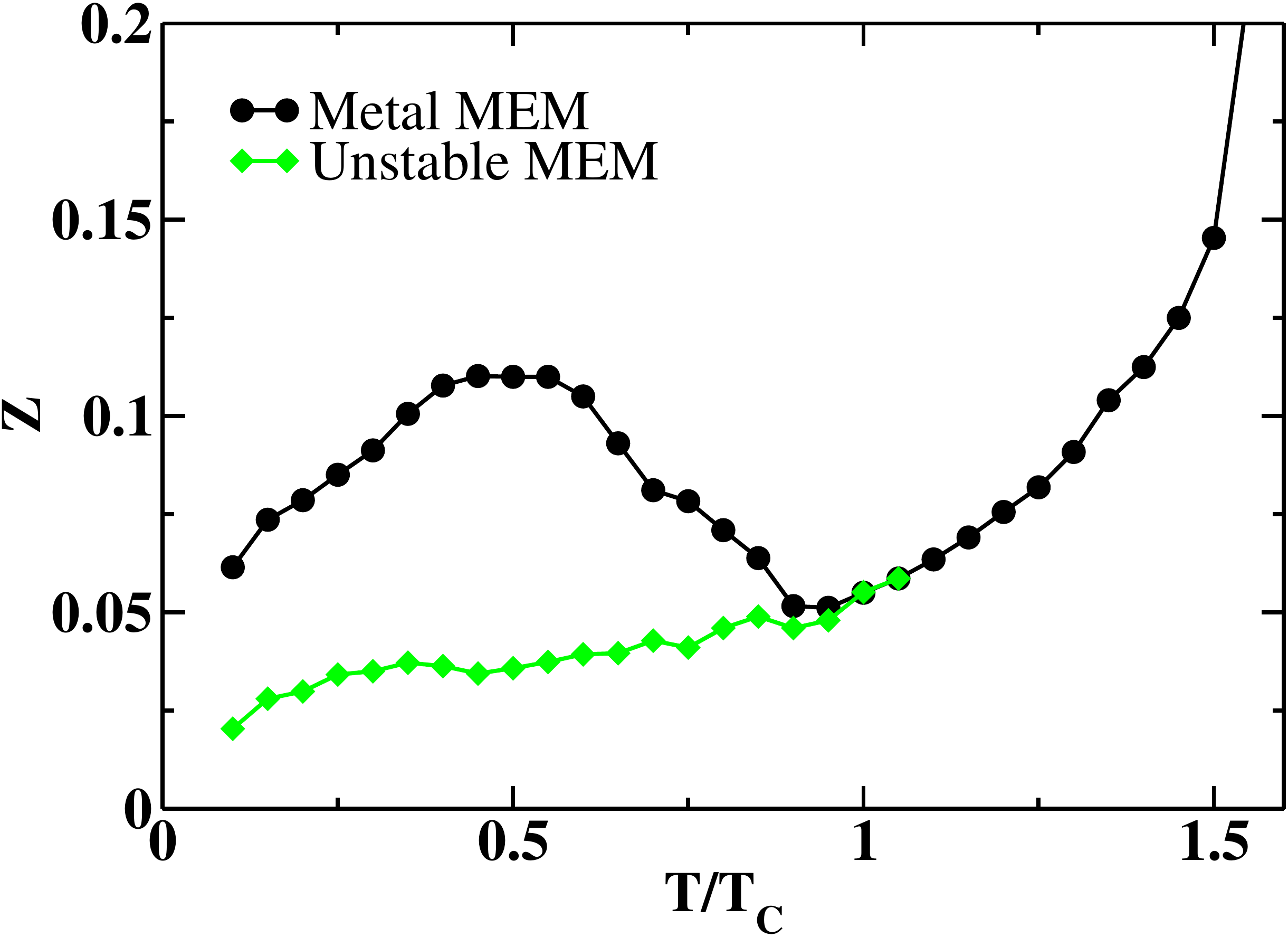}
}\subfloat[CTQMC]{\includegraphics[clip,scale=0.17]{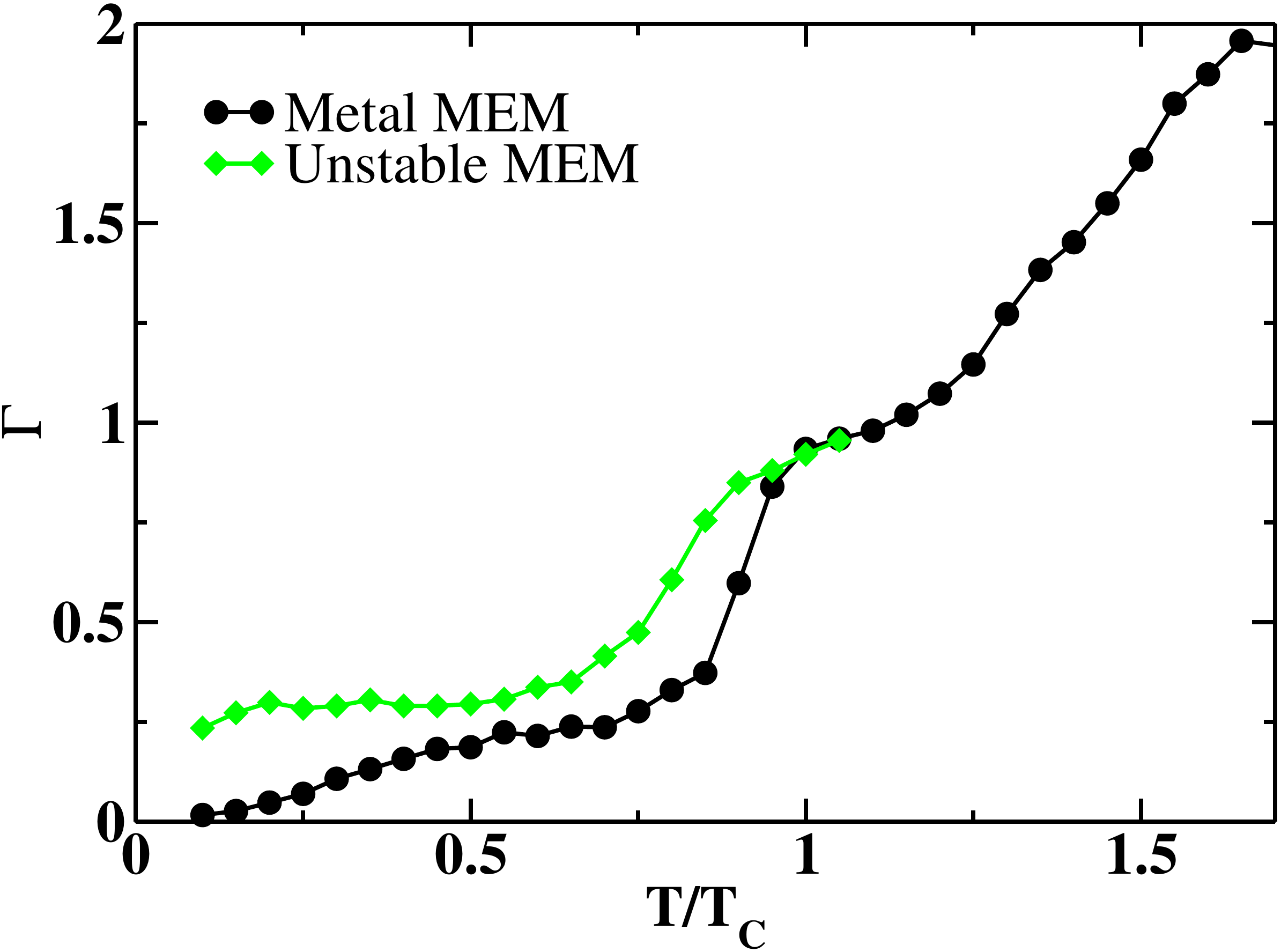}
}
\caption{The resistivity for the metallic (black circles), insulating (red squares), and
unstable (green squares) solutions along the first-order transition
line in a logarithmic $\rho$ scale for (a) CTQMC and (b) IPT.
The Sommerfeld approximations to the resistivities (Eq.~(\ref{eq:sommerfeld})) are shown in the corresponding dashed and dotted lines with different analytic continuation methods (MEM for CTQMC and Pad\'{e}  approximant for IPT).
(c) The quasiparticle weight, $Z$, for the metallic and unstable solutions analytic continued
by MEM (black and green symbols). (d) The scattering rate  $\Gamma$, for the metallic and the unstable solutions, obtained from  MEM (black and green symbols).\label{fig:1st_order_line}}
\end{figure}

\textit{Resistivity calculations and quasiparticle transport.---} To
study the transport properties, we utilize the
Kubo formula \cite{DengRQP} for the DMFT-based DC conductivity
\begin{equation}
\sigma=\sigma_{0}\int d\epsilon\Phi(\omega)\int d\omega(-\frac{\partial f(\omega)}{\partial\omega})A(\epsilon,\omega)^{2},\label{eq:Kubo}
\end{equation}
where the spectral function $A(\epsilon,\omega)=-(1/\pi)\mathrm{Im}(\omega+\mu-\epsilon-\Sigma(\omega))^{-1}$,
$\Phi(\omega)=\Phi(0)[1-(\omega/D)^{2}]^{3/2}$, $\sigma_{0}=2\pi e^{2}/\hslash$, and $f(\omega)$ is the Fermi distribution function.  Here $\Sigma(\omega)$ is obtained on the real axis using standard maximum entropy methods (MEM)
for CTQMC and the Pad\'{e} approximant for IPT. The resistivity is $\rho=1/\sigma$.
To normalize the resistivity, we use the Mott-Ioffe-Regel (MIR) limit $\rho_{MIR}=\hslash D/e^{2}\Phi(0)$, which represent the scale
where the scattering process becomes incoherent and the mean free path is comparable to the Fermi wavelength.

The resistivity calculation, based on Eq.~(\ref{eq:Kubo}), dramatically simplifies in the {\em quasiparticle regime} 
\cite{WenhuHFL}, where transport is dominated by only the leading low-energy excitations. Here the 
Green's function can be approximated as  $G(\omega,\epsilon)\simeq\frac{Z}{\omega-Z\epsilon+i\Gamma_{QP}}$,
with the quasiparticle weight $Z=(1-\frac{\partial \mathrm{Re}\Sigma(\omega)}{\partial\omega})_{\omega=0}^{-1}$
and scattering rate $\Gamma_{QP}=-Z\mathrm{Im}\Sigma(\omega=0)$. A further Sommerfeld approximation 
can be performed in situations where  $\Gamma_{QP}<T$
and $T<ZD$, and the conductivity can be expressed in terms of only two parameters: 
$Z$ and $\Gamma=\Gamma_{QP}/Z$ viz.
\begin{equation}
\frac{\sigma}{\sigma_{MIR}}\approx\frac{1}{\Gamma}\tanh(\frac{Z}{2T}).\label{eq:sommerfeld}
\end{equation}
We explicitly checked that these conditions are indeed obeyed throughout the coexistence region, not only by our metallic solution, but also by the  
unstable solution. As we explicitly show below, the results obtained from numerically evaluating 
the conductivity using our full DMFT solution and Eq.~(\ref{eq:Kubo}) demonstrate remarkable qualitative and even semi-quantitative agreement with our QP approximation. This important result demonstrates the ``resilient'' quasiparticle character  \cite{DengRQP} of transport even for our unstable solution, despite the very unusual behavior of the QP parameters in question. 

\textit{Following the first order transition line.---} To investigate how our three solutions
evolve when approaching the zero-temperature critical point $U_{c2}(T=0)$,
we study the transport properties along the first-order transition
line connecting the two critical
points, $U_{c}(T=T_{c})$ and $U_{c2}(T=0)$.  In Fig.\ref{fig:1st_order_line}(a)
and (b) we show the results obtained from 
CTQMC and IPT, respectively.  At the critical end point $T=T_{c}$, the 
three solutions merge as expected, and the resistivity is of the order
of the MIR limit. Below $T_{c}$, the three solutions trifurcate into different 
trajectories. The unstable solution (green diamonds) displays  higher resistivity
than the metallic solution (black circles), with values of the order
of the MIR limit, $\rho\sim\rho_{MIR}$, indicating bad
metal  behavior \cite{MIR_limit}. The insulating solution has much higher resistivity
due to standard activated transport.

Close to the zero-temperature critical point $U_{c2}(T=0)$   the three solutions 
do not merge, in contrast to the situation around the critical end point ($T\sim T_{c}$).
Instead, while the resistivity of the metallic solution drops at low 
temperatures, that of the unstable solution remains comparable to 
the MIR limit, suggesting incoherent transport, despite these trends being well
captured by the QP approximation (see Fig.~\ref{fig:1st_order_line}(a) and (b)). 
%
%
To better understand this behavior, in 
Fig.~\ref{fig:1st_order_line}(c) we show the quasiparticle weight, $Z$,
for the unstable (green diamonds) and metallic (black circles) solutions. The unstable solution
has significantly  lower $Z$ than the metal, displaying a 
more pronounced  decrease at low temperatures, reminiscent 
of ``resilient'' quasiparticles \cite{DengRQP}. Note that the increase followed by the decrease of $Z$ in the metallic solution may be attributed to the position of the first order transition line in the coexistence region, which shifts towards the insulating phase with lowering the temperature. 
Similar behavior is seen in 
Fig.~\ref{fig:1st_order_line}(d), where we display the behavior of the scattering rate $\Gamma$, which for the unstable solution (green diamonds) remains appreciable down to
the lowest temperatures, again signaling poorly developed (``resilient'') quasiparticles. Remarkably, such non-Fermi Liquid (NFL) behavior 
here persists down to the lowest temperatures, in contrast to previously identified examples of RQPs \cite{DengRQP}, which emerged only at temperatures intermediate between the conventional Fermi Liquid metal at the lowest $T$ and a fully incoherent conductor at high $T$. 

\textit{Constant }$U$\textit{ trajectory.---} In order to further understand
the behavior of the unstable solution in the entire coexistence regime,
we also study the resistivity as a function of $T$, along 
\begin{figure}[t]
\subfloat[CTQMC]{\begin{centering}
\includegraphics[clip,scale=0.163]{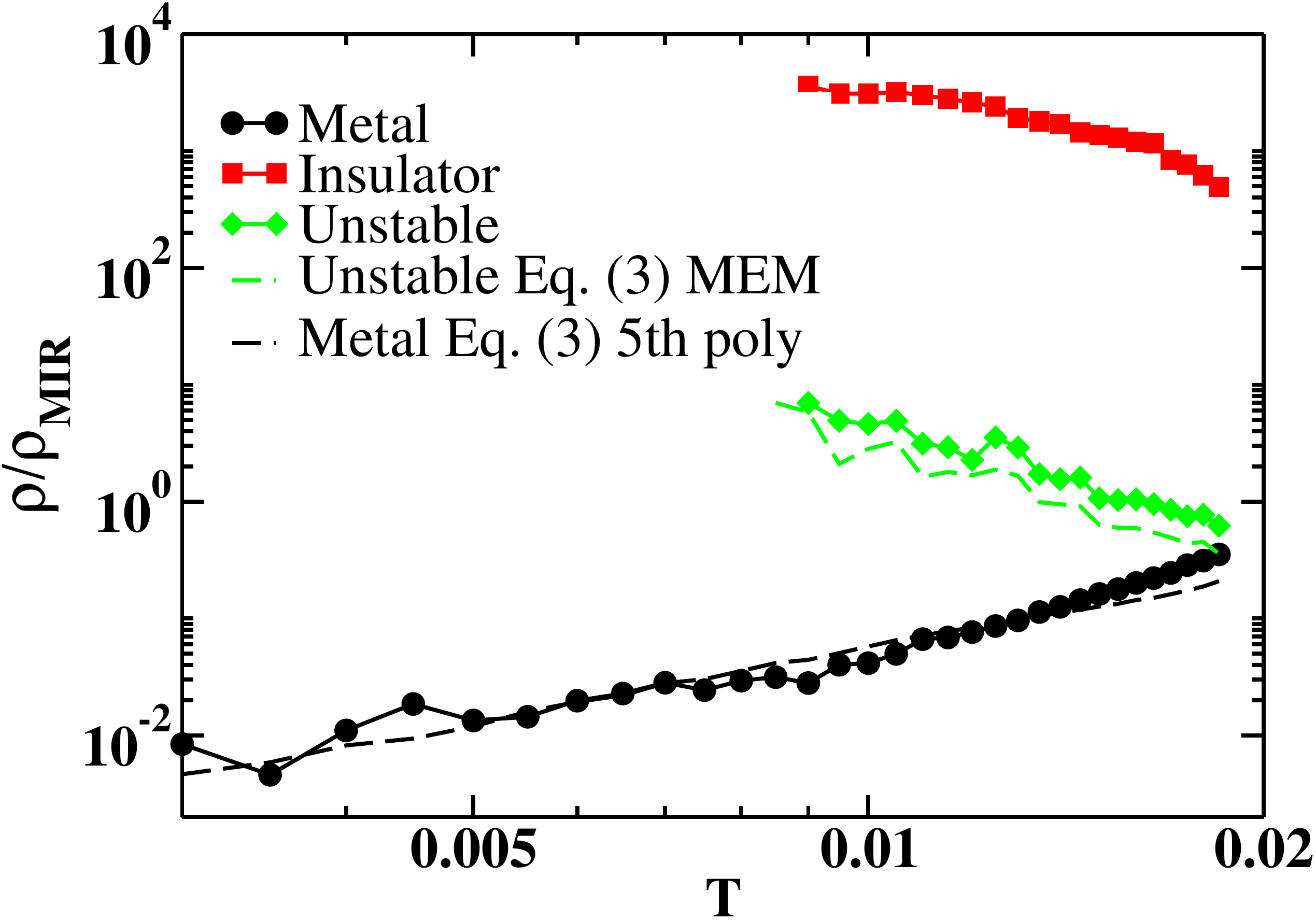}
\par\end{centering}
}\subfloat[IPT]{\begin{centering}
\includegraphics[clip,scale=0.163]{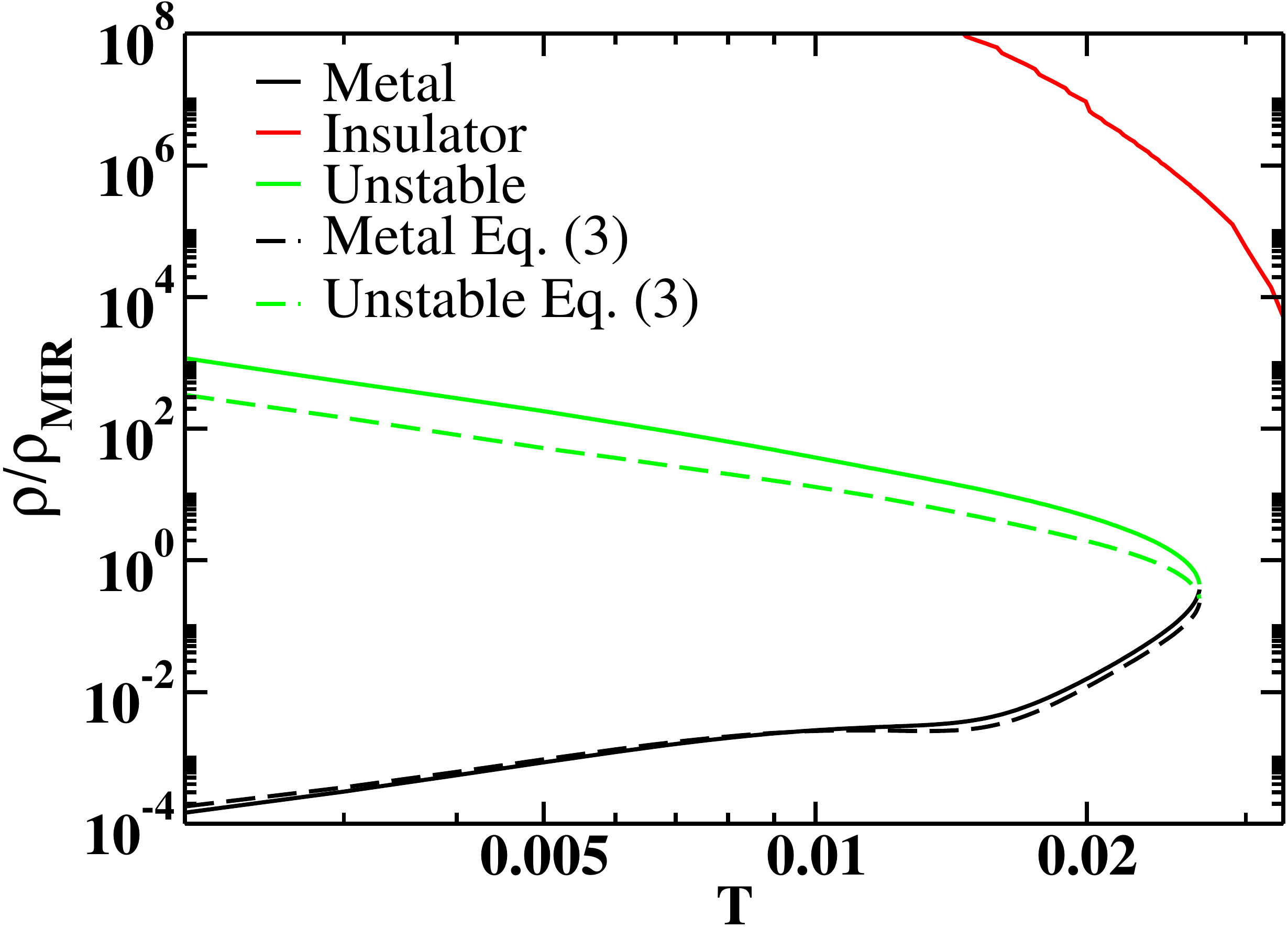}
\par\end{centering}
}

\subfloat[CTQMC]{\begin{centering}
\includegraphics[clip,scale=0.165]{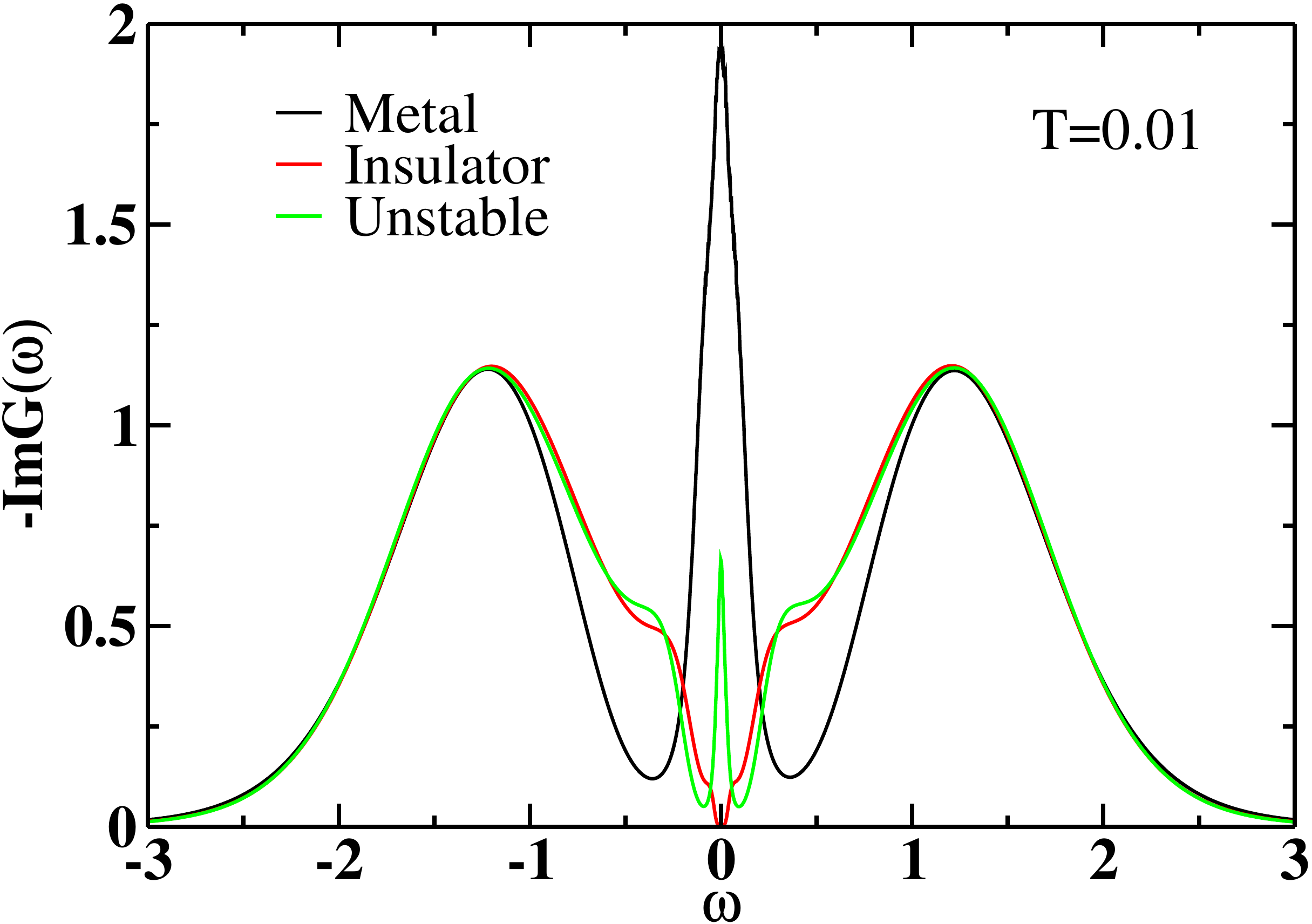}
\par\end{centering}
}\subfloat[CTQMC]{\begin{centering}
\includegraphics[clip,scale=0.165]{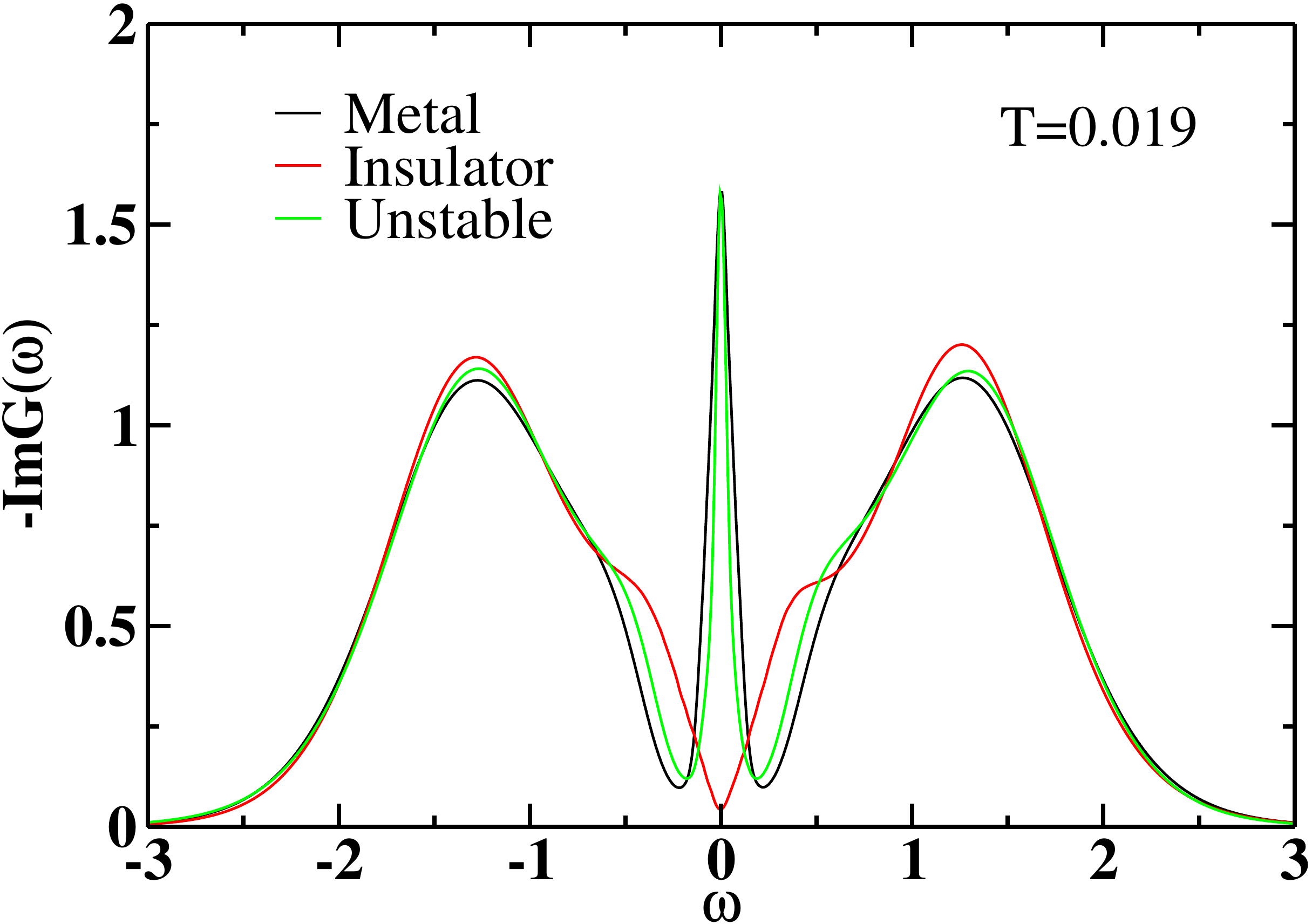}
\par\end{centering}
}

\subfloat[CTQMC]{\begin{centering}
\includegraphics[clip,scale=0.165]{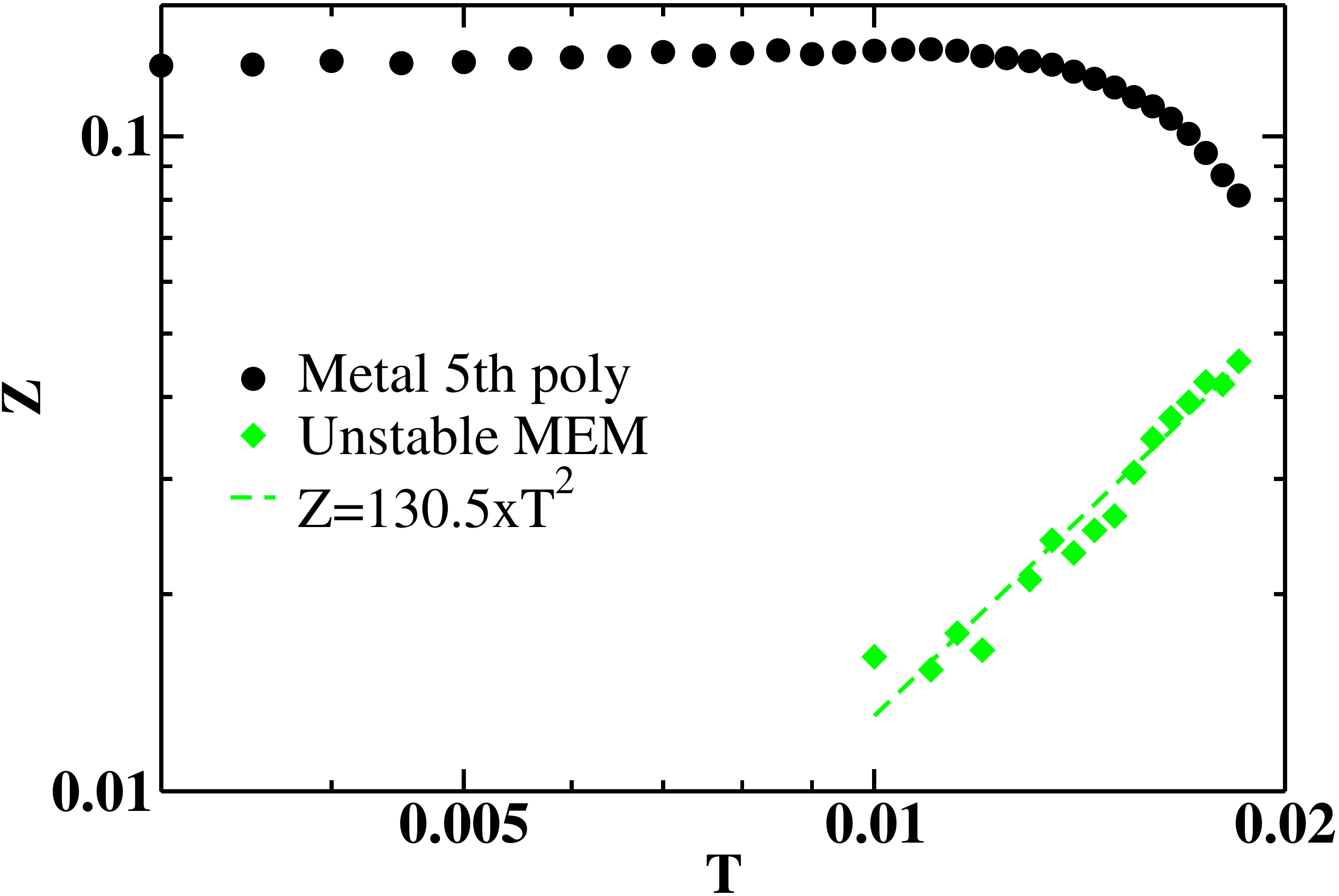}
\par\end{centering}
}\subfloat[CTQMC]{\begin{centering}
\includegraphics[clip,scale=0.165]{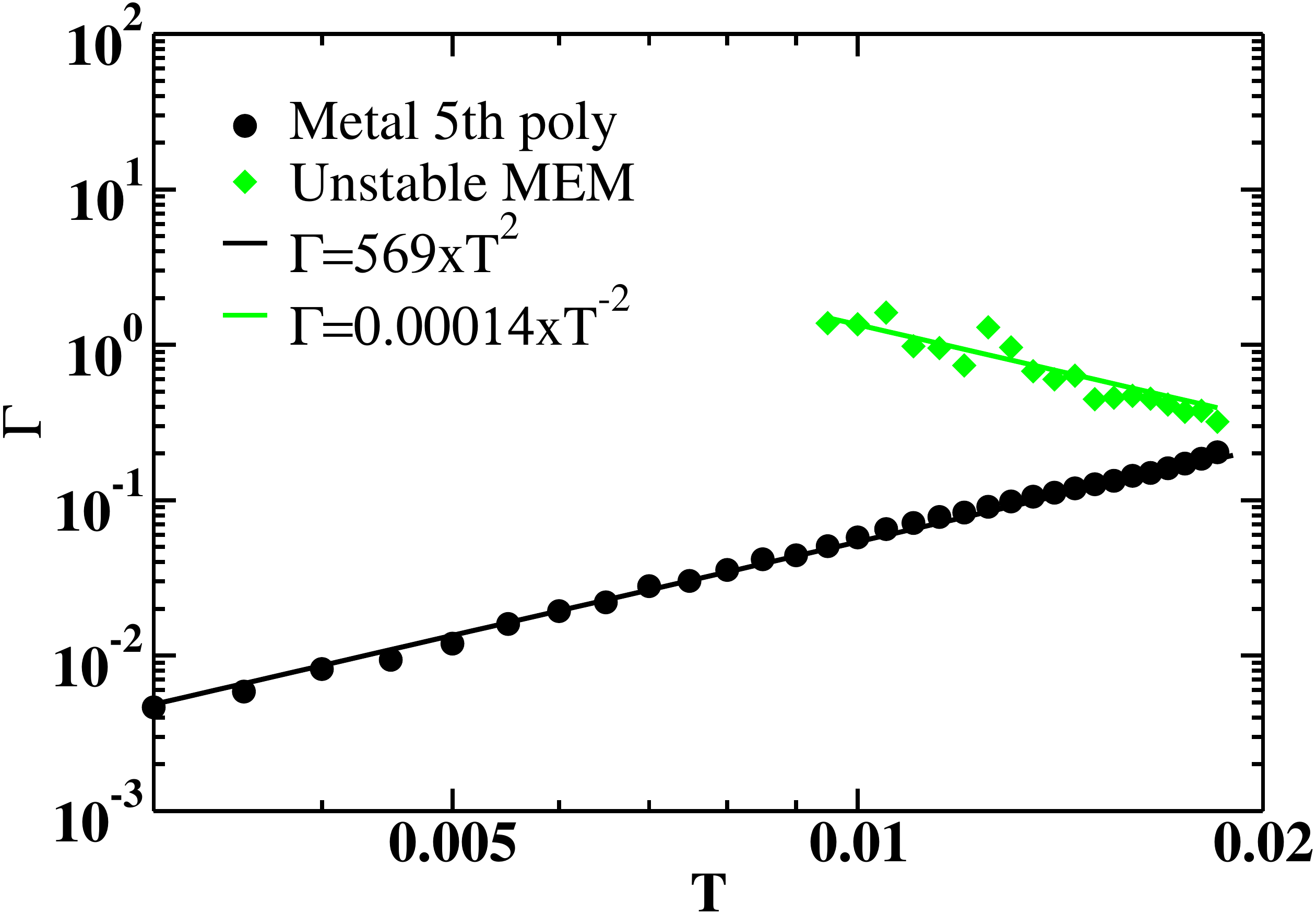}
\par\end{centering}
}
\noindent \centering{}\caption{The resistivity for the metallic (black circles), insulating (red squares), and
unstable (green diamonds) solutions along the constant $U$ trajectory
in logarithmic $\rho$ scale for (a) CTQMC and (b) IPT. The Sommerfeld approximations to the resistivities [Eq.~(\ref{eq:sommerfeld})] are shown in the corresponding dashed and dotted lines with different analytic continuation methods (MEM and polynomial fitting for CTQMC and Pad\'{e} approximant for IPT). The density
of states at  low temperatures (c) $T=0.01$ and (d) $T=0.019$ near the spinodal
line $U_{c2}$ computed from the CTQMC impurity solver and MEM analytic continuation. (e) The quasiparticle weight, $Z$, for the metallic and the unstable
solutions. (f) The scattering rate $\Gamma$ for the metallic and the unstable
solutions.\label{fig:constant_U}}
\end{figure}
\noindent a constant $U$ trajectory. From
the free energy analysis, Fig. \ref{fig:Free_energy}(b), we anticipate that 
the unstable solution gradually shifts towards the insulator as $T$ is reduced at fixed $U$,
eventually merging with it at $T=0$. 
In contrast, in Figs.~\ref{fig:constant_U}(a) and (b), we observe the metallic and insulating solutions displaying 
conventional Fermi liquid and activated behaviors, respectively. On the other hand,
for the unstable solution, we find the resistivity (green diamonds) increases
as the temperature decreases, reaching values as much as two orders of magnitude 
larger than the MIR limit. Nevertheless, as shown in Fig.~\ref{fig:constant_U}(c) and (d),
we observe the unstable solution's density of states (DOS) at the lowest temperature still
features a very small quasiparticle peak at the Fermi level. This suggests 
the unstable solution still retains some metallic character, even though
the resistivity is much larger than the MIR limit. In some sense, this situation could be characterized 
as an extreme example of Bad Metal (BM) behavior  \cite{MIR_limit}, albeit in a  setup which is
dramatically different than the familiar high-$T$ BM behavior in correlated matter. And indeed, the 
standard RQP-Sommerfeld approximation again captures remarkably well all the transport trends, 
even in this extreme high-resistivity regime. 

To even more precisely characterize such RQP-NFL behavior, we next examine the 
corresponding QP parameters, and their evolution as a function of $T$.
In Fig.~\ref{fig:constant_U} (e) we show the quasiparticle weight for $Z$
for metallic (black circles) and unstable (green diamonds) solutions. The metal has a normal behavior with $Z$ saturating at low
temperatures, consistent with the expected FL behavior.
In dramatic contrast,
$Z$ corresponding to the unstable solution decreases rapidly with temperature,
displaying  power-law behavior $Z\sim T^{2}$. Remarkably, a similar but much weaker 
decrease of the form  $Z \sim 1/|\log T|$, dubbed a ``Marginal Fermi Liquid'' (MFL) \cite{MFL98prl}, was proposed
as the key signature of the breakdown of Fermi Liquid theory in optimally doped cuprates. The behavior 
found here is not even ``marginal''. By analogy, it can be described as ``fully developed NFL'' behavior, 
the like of which is seldom seen in correlated matter. Analogously, the unstable solution's scattering rate {\em increases} at low temperatures [Fig.~\ref{fig:constant_U}(f)], again in power-law fashion $\Gamma\sim 1/T^{2}$, well exceeding the MIR limit and consistent with transport.

\textit{Conclusions.---} In this paper, we identified what we argue is a new state of correlated electronic matter, characterizing the domain walls within the metal-insulator coexistence region around the Mott point. We showed that its low-energy excitations display a number of unusual properties, which qualitatively differ from either a conventional metal or an insulator. This paints a physical picture of exotic quasiparticles barely persisting at the brink of an insulating state. Conceptually, such non-Fermi liquid behavior can be viewed as reflecting the quantum critical fluctuations associated with the metal-insulator transition region. 

Our solution was obtained within the framework of single-site DMFT theory, which physically represents the limit of large frustration, where all possible symmetry-breaking fluctuations are suppressed. The MIT coexistence region, however, is presently known \cite{pustogow2021npjQM} to persist in real physical systems such as quasi-two dimensional Mott organics, where inter-site spin correlations could also play a role \cite{pustogow2021science}. The effects of such perturbations can be systematically studied within cluster-DMFT theories \cite{FEVladPRB,VCAunstable}, with indications that the phase coexistence region can be significantly influenced. Nevertheless, we expect the short-range correlation effects will only modify quantitatively the behavior of the unstable solutions revealed in this work. Interesting modifications can also arise by introducing extrinsic disorder due to impurities and defects, which in some cases can significantly reduce the size of the entire coexistence region \cite{kanoda2019prb,kanoda2020prl}. How these perturbations will affect the stability and the relevance of the {\em Domain Wall Matter} we discussed here is a fascinating open problem, which remains a challenge for future experiments as well as theory. 
\section*{acknowledgments}

The work was supported in Brazil
by CNPq through Grant No. 307041/2017-4 and 
Capes through grant 0899/2018 (E.M.). Work in Florida (V. D. and T.-H. L.) 
was supported by the NSF Grant No. 1822258, and the National High Magnetic Field Laboratory 
through the NSF Cooperative Agreement No. 1157490 and the State of Florida. J. V. and D. T. acknowledge funding
provided by the Institute of Physics Belgrade, through the
grant by the Ministry of Education, Science, and
Technological Development of the Republic of Serbia. Numerical simulations were performed on the PARADOX
supercomputing facility at the Scientific Computing
Laboratory, National Center of Excellence for the Study
of Complex Systems, Institute of Physics Belgrade.

\bibliographystyle{apsrev}
\bibliography{unstable} 

\ifarXiv
    

\section*{Supplementary material}

# Supplemental Material: Mott domain walls: a (strongly) non-Fermi liquid state of matter

Tsung-Han Lee,[1,2] J. Vučičević,[3] D. Tanasković,[3] E. Miranda,[4] and V. Dobrosavljević[1]

[1]*Department of Physics and National High Magnetic Field Laboratory,
Florida State University, Tallahassee, Florida 32306, USA*
[2]*Physics and Astronomy Department, Rutgers University, Piscataway, New Jersey 08854, USA*
[3]*Institute of Physics Belgrade, University of Belgrade, Pregrevica 118, 11080 Belgrade, Serbia*
[4]*Gleb Wataghin Physics Institute, The University of Campinas,
Rua Sérgio Buarque de Holanda, 777, CEP 13083-859, Campinas, Brazil*

In this Supplemental Material, we review, in Sec. I, the Ginzburg-Landau theory and the $\phi^4$ theory for dynamical mean-field theory (DMFT) following Refs. [1–3] and the domain-wall solutions in the inhomogeneous Ginzburg-Landau theory. In Sec. II, we provide a detailed comparison of different analytic continuation results for the quasiparticle parameters and transport properties.

## I. GINZBURG-LANDAU THEORY FOR DYNAMICAL MEAN-FIELD THEORY

We start from the Ginzburg-Landau free energy functional for DMFT [1]

$$F_{\text{GL}}[G] = F_{\text{imp}}[G] - \frac{t^2 T}{2}\sum_n G^2(i\omega_n). \tag{1}$$

Then, we expand the free energy functional around the critical point $\delta G = G - G_{\text{cr}}$ leading to

$$F_{\text{GL}} = F_{\text{GL}}^{(0)} + F_{\text{GL}}^{(1)} + F_{\text{GL}}^{(2)} + F_{\text{GL}}^{(3)} + F_{\text{GL}}^{(4)}, \tag{2}$$

where

$$F_{\text{GL}}^{(0)} = F_{\text{GL}}[G_{\text{cr}}], \tag{3}$$

$$F_{\text{GL}}^{(l)} = \frac{T}{l}\sum_{k_1,\ldots,k_l}\Gamma_{k_1,\ldots,k_l}^{(l)}[G_{\text{cr}}]\prod_{j=1}^{l}\delta G_{k_j}, \tag{4}$$

$$\Gamma_{k_1,\ldots,k_l}^{(l)}[G] = \frac{1}{T}\frac{l}{l!}\left(\prod_{j=1}^{l}\frac{\delta}{\delta G_{k_j}}\right)F_{\text{GL}}[G], \tag{5}$$

$k \equiv (\mathbf{k}, i\omega_n)$ is the momentum and frequency index, and $F_{\text{GL}}^{(1)} = 0$ from the DMFT self-consistency condition.

### A. $\phi^4$-model for dynamical mean-field theory

The DMFT Ginzburg-Landau functional can be mapped to a $\phi^4$-model around the critical point [2, 3]. We write

$$\sum_{k_2}\Gamma_{k_1 k_2}^{(2)}[G_{\text{cr}}]\Psi_{\lambda,k_2 p} = \Psi_{\lambda,pk_1}\gamma_{\lambda,p}^{(2)}, \tag{6}$$

where $\Psi_{\lambda,k_1 p}$ and $\gamma_{\lambda,p}^{(2)}$ is the eigenbasis and the eigenvalues of $\Gamma_{k_1 k_2}^{(2)}$, respectively, and $\lambda$ labels the eigenmodes. We then introduce the $\phi_{\lambda,p}$ defined by

$$\delta G_k = \sum_{\lambda,p}\Psi_{\lambda,kp}\phi_{\lambda,p}, \tag{7}$$

such that the Ginzburg-Landau free energy can be written in the eigenbasis as

$$F_{\text{GL}}^{(2)} = \frac{T}{2} \sum_{P} \gamma_P^{(2)} |\phi_P|^2 \tag{8}$$

$$F_{\text{GL}}^{(3)} = \frac{T}{3} \sum_{P_1 P_2 P_3} \gamma_{P_1 P_2 P_3}^{(3)} \phi_{P_1} \phi_{P_2} \phi_{P_3} \tag{9}$$

$$F_{\text{GL}}^{(4)} = \frac{T}{4} \sum_{P_1 \ldots P_4} \gamma_{P_1 P_2 P_3 P_4}^{(4)} \phi_{P_1} \phi_{P_2} \phi_{P_3} \phi_{P_4}, \tag{10}$$

where $P \equiv (\lambda, p)$ and

$$\gamma_{P_1 \ldots P_l}^{(l)} = \sum_{k_1 \ldots k_l} \Gamma_{k_1 \ldots k_l}^{(l)} [G_{\text{cr}}] \prod_{j=1}^{l} \Psi_{\lambda_j, k_j p_j}. \tag{11}$$

As pointed out in Refs. [1–3], the critical behavior is dominated by the soft eigenmode so we can project the free energy functional to this specific eigenbasis. Focusing on the static and uniform part of the free energy function, we can write down the $\phi^4$-model for DMFT

$$\frac{F_{\text{GL}}}{T} = \frac{1}{2} \sum_{p} \gamma^{(2)} \phi_p^2 + \frac{1}{3} \sum_{p_1 p_2 p_3} \gamma^{(3)} \phi_{p_1} \phi_{p_2} \phi_{p_3} + \frac{1}{4} \sum_{p_1 p_2 p_3 p_4} \gamma^{(4)} \phi_{p_1} \phi_{p_2} \phi_{p_3} \phi_{p_4}. \tag{12}$$

### B. Domain-wall solution

To interpret the domain-wall solution, we write the Ginzburg-Landau free energy in the real space

$$\frac{F_{\text{GL}}[\phi]}{T} = \int d\mathbf{x} \left\{ \frac{\kappa}{2} (\nabla \phi(\mathbf{x}))^2 + \frac{1}{2} \gamma^{(2)} \phi(\mathbf{x})^2 + \frac{\gamma^{(4)}}{4} \phi(\mathbf{x})^4 \right\}, \tag{13}$$

where the cubic term in Eq. 12 can be eliminated by properly shifting the field $\phi$ [2, 3]. The Euler-Lagrange equation of $F_{\text{GL}}$ is

$$\gamma^{(2)} \phi(\mathbf{x}) + \gamma^{(4)} \phi(\mathbf{x})^3 - \kappa \nabla^2 \phi(\mathbf{x}) = 0. \tag{14}$$

For simplicity we consider the domain-wall forms along the x-direction, so the Euler-Lagrange equation becomes

$$\gamma^{(2)} \phi(x) + \gamma^{(4)} \phi(x)^3 - \kappa \phi''(x) = 0. \tag{15}$$

with the boundary condition $\phi(x \to -\infty) = -\phi_0$ and $\phi(x \to \infty) = \phi_0$ and $\phi_0 = \sqrt{-\gamma^{(2)}/\gamma^{(4)}}$. Note that the uniform saddle-point solution $\phi_0$ can be related to the double occupancy or the density of states at the Fermi level, where the positive and negative values of $\phi_0$ correspond to the metallic and the insulating solutions, respectively [2]. One can show that the inhomogeneous domain-wall solution is [4]

$$\phi(x) = \phi_0 \tanh\left[\frac{x}{\sqrt{2}\xi}\right], \tag{16}$$

where $\xi = \sqrt{-\kappa/\gamma^{(2)}}$ is the correlation length. We can see that the domain-wall solution at $\phi(x) = 0$ corresponds to the local maximum of the uniform Landau free energy

$$f[\phi] = \frac{1}{2} \gamma^{(2)} \phi^2 + \frac{\gamma^{(4)}}{4} \phi^4 \tag{17}$$

shown schematically in Fig. 1.

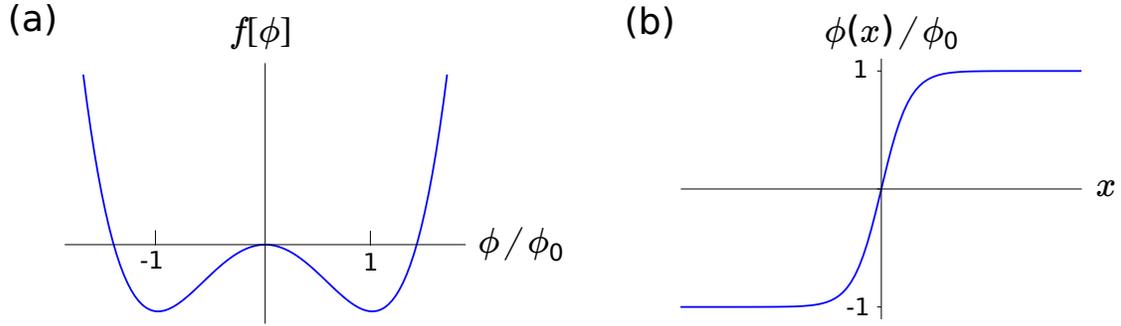

FIG. 1. (a) Schematic Landau free energy $f[\phi]$ as a function of $\phi$. (b) Schematic domain-wall solution profile $\phi(x)$ as a function of $x$.

## II. COMPARISON OF MAXIMUM ENTROPY METHOD AND POLYNOMIAL FITTING ANALYTIC CONTINUATION

In this section, we compare the quasiparticle and transport properties obtained from two analytic continuation approaches: maximum entropy method (MEM) and fifth-order polynomial fitting.

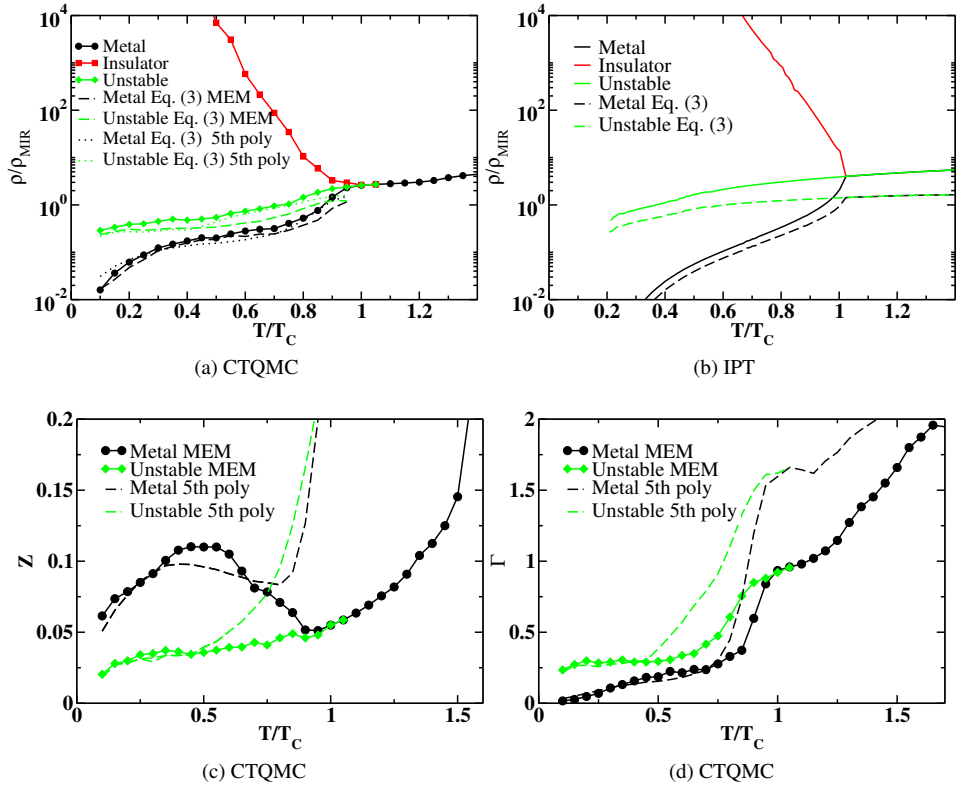

FIG. 2. The resistivity for the metallic (black circles), insulating (red squares), and unstable (green squares) solutions along the first-order transition line in a logarithmic $\rho$ scale for (a) CTQMC and (b) IPT. The Sommerfeld approximations to the resistivities (Eq. (3)) are shown in the corresponding dashed and dotted lines with different analytic continuation methods (MEM and polynomial fitting for CTQMC and Padé approximation for IPT). (c) The quasiparticle weight, $Z$, for the metallic and unstable solutions analytic continued by MEM (black and green symbols) and by polynomial fitting (black and green dashed lines). (d) The scattering rate $\Gamma$, for the metallic and the unstable solutions, obtained from MEM (black and green symbols) and from polynomial fitting (black and green dashed lines).

First, we discuss the results along the first-order transition line. Figure 2(a) shows the resistivities computed from the Som-

merfeld approximation (Eq. (3) in the main text) using MEM and polynomial fitting analytic continuations. We see that the two analytic continuations agree at low temperatures. However, around $T_c$, the two methods show significant differences. This behavior is also observed in the quasiparticle weight $Z$ (Fig. 2(c)) and the scattering rate $\Gamma$ ((Fig. 2(d)). The difference between the two analytic continuation methods stems from the breakdown of the polynomial fitting around $T_c$, where the polynomial fitting of the first few Matsubara points of the self-energy does not yield reliable quasiparticle parameters $Z$ and $\Gamma$. Therefore, around $T_c$, one should trust the MEM results. On the other hand, the polynomial fitting should be more reliable at low temperatures, where the first few Matsubara points are enough to determine the quasiparticle parameters, whereas the MEM is noisier.

We now discuss the results of different analytic continuation approaches along a constant $U$ trajectory. In Fig. 3(a), we show the resistivity calculated from the Sommerfeld approximation (Eq. (3) in the main text) using MEM and polynomial fitting analytic continuation. We observe that the two analytic continuation approaches give semi-quantitatively similar behavior. In Fig. 3(c), we show the quasiparticle weight $Z$ calculated from the two analytic continuation approaches. We observe that the two analytic continuations give the same behavior for the metallic solutions quantitatively. Note that the polynomial fitting is expected to give a more reliable analytic continued $Z$ at low temperature. On the other hand, for the unstable solutions, we see that the two analytic continuation approaches yield different powers where the polynomial fitting gives $Z \sim T^{1.5}$ and MEM gives $Z \sim T^2$. Both differ significantly from the Fermi-liquid saturation behavior shown in the metallic solutions. Finally, in Fig. 3(d), we show the scattering rate $\Gamma$ calculated from the two analytic continuation approaches. We observed that the two approaches give semi-quantitatively the same behavior with similar power, $\Gamma \sim T^2$ for the metallic solutions and $\Gamma \sim T^{-2}$ for the unstable solutions. We note that the polynomial fitting is expected to yield more reliable analytic continued $\Gamma$ for the metallic solutions at low temperatures.

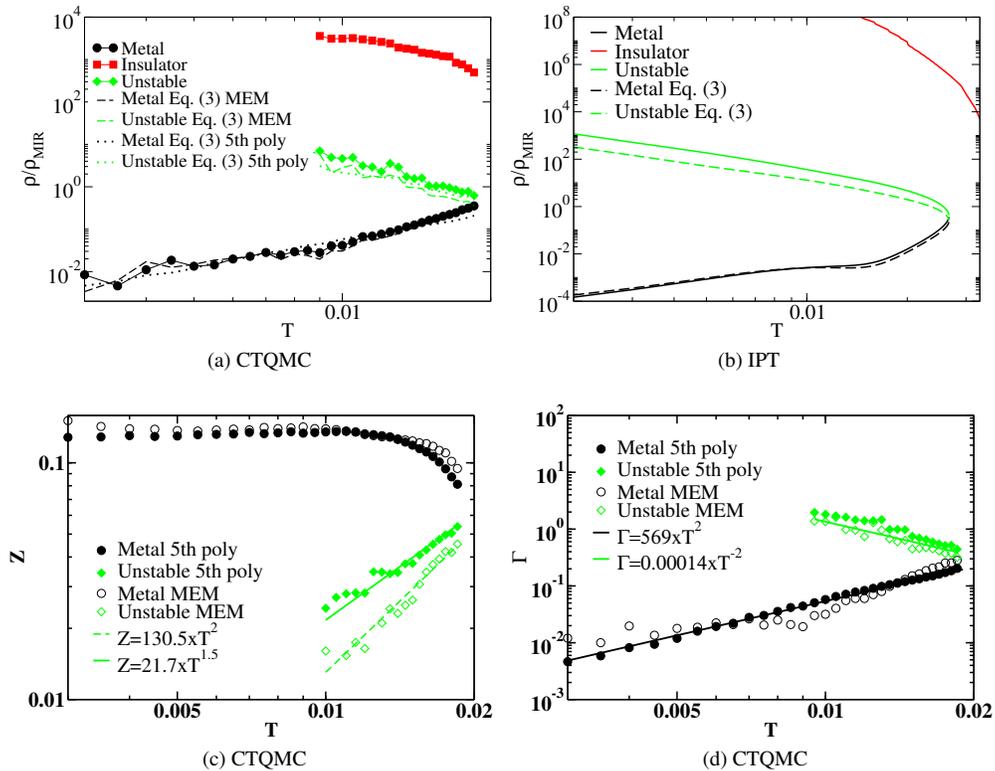

FIG. 3. The resistivity for the metallic (black circles), insulating (red squares), and unstable (green diamonds) solutions along the constant $U$ trajectory in logarithmic $\rho$ scale for (a) CTQMC and (b) IPT. The Sommerfeld approximations to the resistivities are shown in the corresponding dashed and dotted lines with different analytic continuation methods (MEM and polynomial fitting for CTQMC and Padé approximation for IPT). (c) The quasiparticle weight, $Z$, for the metallic and the unstable solutions. (d) The scattering rate $\Gamma$ for the metallic and the unstable solutions.


\fi

\end{document}